\documentclass[runningheads]{llncs}
\usepackage{graphicx}
\usepackage{todonotes}
\usepackage{booktabs}
\usepackage{makecell}
\begin{document}
\title{\textit{That's All Folks}: a KG of Values as Commonsense Social Norms and Behaviors}
%
%
\author{Stefano De Giorgis\inst{1}\orcidID{0000-0003-4133-3445} \newline \and 
Aldo Gangemi\inst{1,2}\orcidID{0000-0001-5568-2684}}
\authorrunning{De Giorgis \& Gangemi}
%
\institute{University of Bologna, Via Zamboni 32, 40126 Bologna (BO), Italy \and
ISTC - CNR, Via S. Martino della Battaglia 44, 00185 Roma (RM), Italy
\email{\{stefano.degiorgis2,aldo.gangemi\}@unibo.it}\\
}
\maketitle              
\begin{abstract}
Values, as intended in ethics, determine the shape and validity of moral and social norms, grounding our everyday individual and community behavior on commonsense knowledge. 
Formalising latent moral content in human interaction is an appealing perspective that would enable a deeper understanding of both social dynamics and individual cognitive and behavioral dimension.
To tackle this problem, several theoretical frameworks offer different values models, and organize them into different taxonomies. The problem of the most used theories is that they adopt a cultural-independent perspective while many entities that are considered ``values'' are grounded in commonsense knowledge and expressed in everyday life interaction.
We propose here two ontological modules, \textit{FOLK}, an ontology for values intended in their broad sense, and \textit{That's All Folks}, a module for lexical and factual folk value triggers, whose purpose is to complement the main theories, providing a method for identifying the values that are not contemplated by the major value theories, but which nonetheless play a key role in daily human interactions, and shape social structures, cultural biases, and personal beliefs.
The resource is tested via performing automatic detection of values from text with a frame-based approach.

\end{abstract}
\section{Introduction }

Moral and social values are considered to be the principles, beliefs, and attitudes that guide our behavior and shape our interactions with others.

Values are important for social interactions and social structures, because they provide a framework for understanding and evaluating human personal actions and societal dynamics of interaction. Moral values such as fairness, justice, and compassion help us to determine what is right and wrong, and they are used as basis for building and maintaining social and emotional connection with others.

Among other theoretical frameworks, Moral Foundations Theory (MFT) \cite{graham2013moral} \cite{graham2012moral} \cite{haidt2012righteous} \cite{haidt2007new} is a theory that proposes  six fundamental moral values as universal across cultures: care, fairness, loyalty, authority, sanctity, and liberty. According to this theory, values are innate and hardwired into the human brain, and they are pillars for  moral reasoning and decision-making. Another well established framework is Basic Human Values (BHV) \cite{schwartz1992universals} \cite{schwartz2001extending} \cite{schwartz2012refining}. In its original version, it proposes ten basic values underlying all human motivations and actions: power, achievement, hedonism, stimulation, self-direction, universalism, benevolence, tradition, conformity, and security.

The ValueNet ontology is an ontology to represent and reason over knowledge about values. The two abovementioned theories are modelled and axiomatised in OWL \cite{de2022basic}, and they are introduced in the ValueNet ontology as modules, and used for answering competency questions about the domain of values. \newline

The ontological modules presented in this work are instead motivated by a more factual and pragmatic approach. In fact, there is a huge scholarly debate about \textit{what} and \textit{how many} moral values should be considered, but people have commonsense knowledge about behaviors that shape everyday social interactions, and are able to answer (or at least to elaborate, if asked) questions like ``what is that you look for in a good friend?'' or ``what do you evaluate the most in your search for a soulmate?''.
Consider for example an elaborated version of the famous trolley dilemma \cite{yang2022differences}: is ``being fit'' a value that can be a \textit{discrimen} between life and death? Does ``having a healthy life'' make an individual deserve to be preferred in a list of people waiting for a vital organ transplant? Is ``being rich'' a value? Answers over the literature and the web widely vary.

Since social values seem to be widespread societal norms and beliefs, in this work we present a preliminary and pragmatic reverse engineering of commonsense knowledge about values.
Our intention is to provide, next to the formalisation of main traditional theories, a module that considers a bottom up, folk-determined perspective, and which allows to detect and reason over culturally dependent (and therefore debatable) entities such as ``fitness'', ``punctuality'', ``wealth'', etc.



We begin with some Competency Questions that the FOLK ontological module intends to answer.

\begin{enumerate}
    \item Is the entity \textit{x} an instance of some folk value, according to commonsense knowledge about ``values''?
    \item What is the relation among Folk Values and BHV or MFT values ?
\end{enumerate}

Furthermore, from the operationalisation of the FOLK module, realised via performing a SPARQL query expansion as explained in Sec. \ref{sec:quokka}, with the \textit{That's All Folks} (TAF) module, we perform a frame-based automatic detection of values, as explained in Sec. \ref{sec:TAF}, in order to answer the following CQs:
\begin{enumerate}
    \item Given a sentence, is there some FV trigger?
    \item What are the semantic relations among the trigger and other elements of the sentence?
    \item What is the epistemic stance towards an entity in a sentence, according to its value-connotation?
    
\end{enumerate}

The paper is organized as follows: Sec. \ref{sec:theoretical_grounding} presents the main theoretical background, in particular Sec. \ref{subsec:BHV_theory} is focused on BHV theory, while Sec. \ref{subsec:MFT_theory} presents MFT main assumptions; Sec. \ref{sec:ValueNet} describes the BHV and MFT ontological modules in the ValueNet ontology; Sec. \ref{sec:quokka} describes briefly the QUOKKA workflow, used to populate the Folk Values ontology knwoledge graphs; the newly introduced FOLK and TAF modules, and their integration in the ValueNet ontology are presented in Sec. \ref{sec:TAF}, while finally the results from resource evaluation are described in Sec. \ref{sec:eval}.

\section{Theoretical Grouding}
\label{sec:theoretical_grounding}

In this Section, we briefly describe the main theories modeled in the ValueNet ontology, pointing out strengths and weaknesses for each of them.

\subsection{Basic Human Values Theory}
\label{subsec:BHV_theory}

The Theory of Basic Human Values (BHV) was proposed by Shalom Schwartz in the 1980s as a way to understand human values in a cross-cultural context. According to this theory, human values are arranged in a "value wheel" with two axes that divide the values into four quadrants. The value wheel model also includes a congruity continuum that connects adjacent values.

\begin{figure}
\label{fig:circumplex_model}
  \centering
  \includegraphics[scale=0.2]{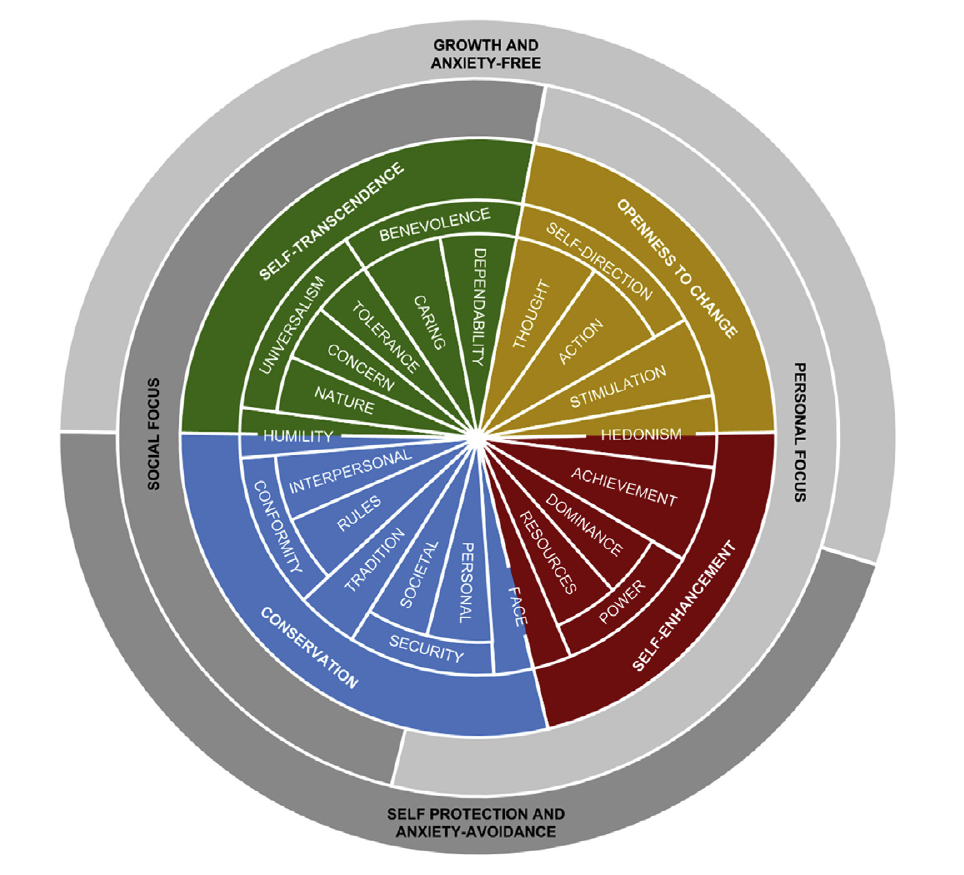}
  \caption{Basic Human Values circumplex model, image taken from [Giménez, August Corrons, and Lluís Garay Tamajón. "Analysis of the third-order structuring of Shalom Schwartz’s theory of basic human values." Heliyon 5.6 (2019): e01797.]}
\end{figure}

In its first version, the model included $10$ values \cite{schwartz1992universals}, but, after testing it in social experiments and some rework the model was later refined to $19$ values in total \cite{schwartz2012refining}, as shown in Fig. $3$.
BHV relies on the opposition and similarity of values, grouped into macro-categories that are mostly determined by (i) individual personality traits (self-transcendence vs self-enhancement, conservation vs openness to change) and (ii) the ``focus'' of a value, namely the main beneficiary target (personal vs societal). This model has inspired the design of a questionnaire (Portrait Values Questionnaire, PTV) which has been employed by a number of studies to explore values across different countries \cite{schwartz2001extending}. In later works \cite{schwartz2012overview}, Schwartz provides evidence in favour of a pan-cultural arrangement of value priorities.

Albeit BHV theory has been tested on a large number of subjects across $82$ countries, a note of criticism that still stands, is its top-down approach, having established the number and taxonomy of values a priori, and then started validating it through dedicated experimentation.

\subsection{Moral Foundations Theory}
\label{subsec:MFT_theory}

The Moral Foundation Theory (MFT) is a theoretical framework whose main strength is to explain its moral and social values as not depending on culture. It is based on the work of Schweder et al. \cite{mahapatra1997divinity} on universal human ethics and the study of moral emotions, particularly the role of behavioural neuro-cognitivism. MFT takes an agnostic approach to cultural influences, recognizing that, while the ontological existence of the value-violation dyadic oppositions, enumerated in the followings, is culture-independent, their realization in real-world situations depends on the specific values that are being considered. Adopting terms from logics: they are universal in their intension, but cultural-determined in their extensional occurrences (e.g. the notion of \textit{sanctity vs degradation} is common to all cultures, but what is actually considered a degradating behavior is subject to socio-cultural, topological and chronological variables). The model proposed by Graham et al. \cite{graham2009liberals} focuses specifically on the opposition of single values, in which any pair of opposing values represents a dyad that either promotes or restricts some behavior. According to MFT, there are six innate moral foundations that can be found across cultures and societies:

\begin{itemize}
    \item \textit{Care vs Harm}: this foundation is grounded in the attachment systems and is related to our concern for the well-being of others. It underlies the notion of empathy, intended as the ability to not only understand, but also feel, the same feelings as others, thus being able to imagine hypothetical scenarios, in which we are living some positive or negative mental or physical state, which we actually don't live.
    \item \textit{Fairness vs Cheating}: is grounded in the evolutionary process of reciprocal altruism, it motivates humans to treat others equally and avoid cheating or taking advantage of them.
    \item \textit{Loyalty vs Betrayal}: is grounded in the clans and family-based dimension that for a long time characterized most of our tribal societies. The ability to create links and alliances was a way to increase the surviving percentage possibilities for oneself and his/her close group, it motivates people to act in ways that support their social group and defend it from harm.
    \item \textit{Authority vs Subversion} is grounded in the hierarchical social interactions directly inherited by primates’ societies it is the foundation motivating respect for social norms in every society.
    \item \textit{Sanctity vs Degradation} is deeply associated with the CAD triad emotions (Contempt, Anger, Disgust) and the psychology of disgust, it is one of the most spread dyadic oppositions, underlying religious (and not only) notions of living in an elevated, less carnal, more ascetic way. It underlies the idea of ``the body as a temple'' which can be contaminated by immoral activities and it is foundational for the opposition between soul and flesh. 
    \item \textit{Liberty vs Oppression} is grounded in feelings and experiences of self - determination, freedom of thought and expression vs. episodes of unjustified violence or liberty restrictions.
\end{itemize}

Besides its relevance for the investigation of the emotional counterpart of value appraisal and for the cross-cultural investigation of values, MFT has inspired the design of the Moral Foundation Dictionary \cite{graham2011mapping} and, more recently, of the Extended Moral Foundations Dictionary \cite{hopp2021extended}, which combine theory-driven elements on moral intuitions with a data-oriented approach.

The most recent resource realized adopting MFT is the Moral Foundation Reddit Corpus \cite{trager2022moral}: a dataset of 35k sentences, taken from many subreds from the Reddit social network, and labeled with single poles (positive or negative) of the MFT dyads.
In this work it is adopted a subset of the MFRC to perform evaluation of the proposed ontological modules.



\section{ValueNet Ontology}
\label{sec:ValueNet}

The ValueNet ontology is the ontology network that aims at representing human moral and cultural values. 

\subsection{Technical Preliminaries}
\label{subsec:preliminaries}

The brief introduction of some resources and technical approaches is necessary to understand the perspective adopted in the presented work.

\paragraph{\textbf{Frame Semantics and Framester}}
Frames are, in a broad definition, cognitive representations of typical features of a situation. Fillmore's frame semantics \cite{fillmore1982framsemantics} has had the largest impact as structuring the combination of linguistic descriptors and features of related knowledge structures to describe cognitive phenomena. Lexical units and sentences are semantically associated with ``frames'', namely schematic structures, based on the common scene they evoke. In FrameNet\cite{nuzzolese2011gathering}, a formal representation of Fillmore's frame semantics, frames are also explained as \textit{situation types}.

Framester ontology hub \cite{gangemi2016framester,gangemi2020closing} provides a formal semantics for frames in a curated linked data version of multiple linguistic resources (e.g. besides FrameNet, WordNet \cite{miller1998wordnet}, VerbNet \cite{schuler2005verbnet}, a cognitive layer including MetaNet\cite{gangemi2018afaoocm} and ImageSchemaNet \cite{de2022imageschemanet} BabelNet \cite{navigli2010babelnet}, etc.), factual knowledge bases (e.g. DBpedia \cite{auer2007dbpedia}, YAGO \cite{suchanek2007yago}, etc.), and ontology schemas (e.g. DOLCE-Zero \cite{gangemi2003sweetening}), with formal links between them, resulting in a strongly connected RDF/OWL knowledge graph.

\subsection{ValueNet Ontology Network}
\label{subsec:modules}

ValueNet is developed as a knowledge layer related to values in the Framester ontology hub, therefore it adopts a frame semantics approach, it reuses entities from the Framester hub, it inherits the usage of ODPs, and it is aligned to the DOLCE foundational ontology.
It exploits the OWL2 punning technique to include double representation of values both intensionally, as conceptual frames, and extensionally, as classes of situations, namely frame occurrences. 

Apart from the ValueCore module, which provides the minimum semantic vocabulary to speak about values as concepts, some theoretical modules developed since now are: 
\begin{itemize}
    \item MFT: is the ontological transposition of the Moral Foundations Theory, as described in the previous Section;
    \item BHV: the formalisation of Schwartz's ``Value Wheel'' as in \cite{gimenez2019analysis}, described in Sec. \ref{subsec:BHV_theory};
    \item CMM: the Moral Molecules Theory, taken from cognitive psychology, as exposed by Curry in \cite{curry2021moral};
    \item MFTriggers: the module that operationalizes the MFT theory and allows the automatic extraction of moral values from natural language;
    \item BHV triggers: parallel to MFTriggers, it operationalizes the BHV theory;
    \item FOLK: the newly realised ontological transposition of the Folk Value lists, scraped from the web, as explained in Sec. \ref{sec:TAF}, then aligned to MFT and BHV;
    \item TAF: (That's All Folks) the module operationalising FOLK, available online permanently both on the Framester resource\footnote{http://etna.istc.cnr.it/framester2/sparql} or on the ValueNet GitHub\footnote{https://github.com/StenDoipanni/ValueNet/tree/main/ThatsAllFolks}.
\end{itemize}

\section{Populating the Folk Value Ontology}
\label{sec:quokka}

\begin{figure}
\label{fig:populate_kg}
  \centering
  \includegraphics[scale=0.2]{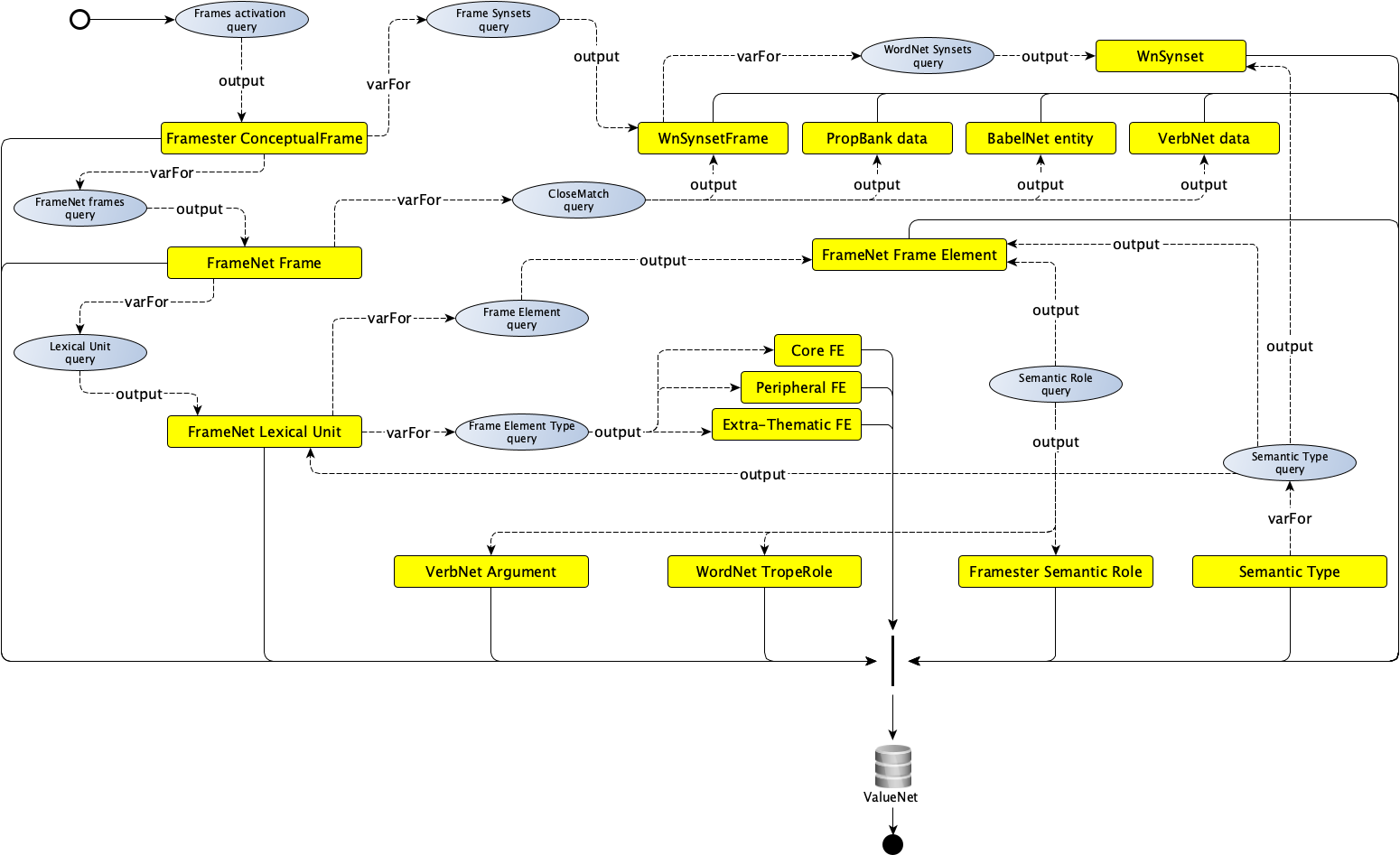}
  \caption{The QUOKKA workflow to semi-automatically populate knowledge graphs, starting from lexical units to entities from semantic web resources aligned in the Framester hub.}
\end{figure}

To operationalise the ontological resources in ValueNet, and to consequently be able to exploit the inference power of the ontological structures, it has been applied the QUOKKA workflow, shown in Fig. \ref{fig:populate_kg}. Here we briefly mention the general rationale and the most productive queries used to populate the TAF resource, but a detailed description and several SPARQL queries are available on the documented repository\footnote{Here you can find the QUOKKA GitHub repository: \url{https://github.com/StenDoipanni/QUOKKA}}.

Fig. \ref{fig:populate_kg} shows how entities activating positive or negative values (in MFT terms: values or violations), represented as rectangular boxes, are retrieved via SPARQL queries to the Framester endpoint, represented as oval shapes and described in the next paragraphs. Fig. \ref{fig:populate_kg} furthermore shows how some entities, being retrieved by some query, are used as input for further queries. Rectangular boxes with no incoming \texttt{output} and only \texttt{:varFor} arrow represent those steps that need a human in the loop (e.g. the semantic type query, to produce meaningful results, requires some domain expert which analyzes all results and filters them manually). Ovals with no \texttt{varFor} incoming arrow represent those queries which need a human in the loop providing some \texttt{input\_variable}. All other steps are automatized, although, due to the great amount of knowledge in the Framester resource, a manual check could result in higher quality data.
Each query can be reproduced on the Framester endpoint\footnote{The Framester endpoint is available at \url{http://etna.istc.cnr.it/framester2/sparql}} by substituting (manually or programmatically) the \texttt{insert\_variable} element with the corresponding entity.

\paragraph{\textbf{Manual Lexical Units Selection}}
To populate the TAF module, the first step consists in taking all the lexical units used to denote values in the scraped lists, and use those as input variables for the workflow. For example, two different lists presented the concept of \texttt{folk:Risk} as a value, providing the following definition ``If you value taking risks, you know that if you follow your gut, there is a chance that it will lead to a huge payoff. You’re not afraid to face the option of failing if there is also an option for success''. 
Therefore, the first step is to use the lexical unit ``risk'' as input variable, to investigate the semantic area related to the notion of \textit{risk}.

\paragraph{\textbf{Frame-driven triggering}}
After the manual selection, the lexical units are used as input variables for querying the Framester resource, looking for frames activating the value/violation. This first SPARQL query uses a non-disambiguated lexical unit (e.g. \textit{risk}) in order to investigate the presence of triggers for the \texttt{folk:Risk} value). For example, for \textit{risk}, the result of the query includes all its possible senses, and their relative evoked frames.
This query is represented in  Figure \ref{fig:populate_kg} as the ``Frames activation query'' starting node. In our case, the frames retrieved are \texttt{fs:RiskySituation}, \texttt{fs:RunRisk}, \texttt{fs:BeingAtRisk}, \texttt{fs:Daring} and \texttt{fs:Endangering}.
After performing the query, the selection of frames possibly triggering some value is done manually and the first phase of frames activation search is closed.

\paragraph{\textbf{Concept-driven triggering}}
Parallel to the Frame-driven activation, non- disambiguated lexical units are used to retrieve entities from ConceptNet, exploiting its semantic relations, in particular are considered the following: \texttt{cn:DerivedFrom}, \texttt{cn:Causes}, \texttt{cn:IsA}, \texttt{cn:UsedFor}, \texttt{cn:HasSubevent} and \texttt{cn:FormOf}.
This query is represented in Fig. \ref{fig:populate_kg} as ``Concept activation query''.

\paragraph{\textbf{DBpedia factual triggering}}
From ConceptNet concepts it is possible to retrieve aligned entities from WikiData and DBpedia, providing a factual grounding to TAF knowledge base.
This query is shown in Fig. \ref{fig:populate_kg} as ``DBpedia External URL query''.
In our case, the \texttt{dbpedia:Risk} entity and the WikiData \texttt{wiki:Q104493} are retrieved, as well as many entries from the wiktionary (e.g. \textit{risky}, \textit{riskful}, \textit{risktaker} etc.).

\paragraph{\textbf{Frame element-driven triggering}}
Frame element activation concerns the activation of some semantic roles related to a ``Value situation'', and can be performed similarly as with frames.
This kind of query is exemplified by focusing on retrieving FrameNet frame elements of type ``Core'', ``Extra-Thematic'' and ``Peripheral''.
The query is shown in Fig. \ref{fig:populate_kg} as Frame Element Type Query.

\paragraph{\textbf{Lexical unit-driven triggering}}
Activation from lexical material is substantial to perform the semantic value detection, described in Sec. \ref{sec:eval} and it is generated automatically re-entering in the workflow the ``Frame activation query'' results, namely the frames previously manually selected.
The rationale is that if some entity evokes a frame, which in turn triggers a value, than that entity should have some form of activation to the value itself.
Therefore this query extracts all the elements (typically WordNet synsets) that evoke a frame. The query is performed for all the frames retrieved and selected as activators by the Frame Activation query.
Thanks to further alignments it is possible to exploit the word-sense-key relation to retrieve entities from VerbNet aligned to WordNet senses, and include them too as value lexical triggers.
In our case, the WordNet \texttt{wn:risk-verb-2}, \texttt{wn:gamble-verb-1} and \texttt{wn:venture-verb-3} synsets are retrieved, and from those, in turn, the \texttt{vn:Risk\_94000000}, \texttt{vn:Gamble\_70000000} and \texttt{vn:Venture\_94100000}, as well as many others, are retrieved.

This query could be found in Fig. \ref{fig:populate_kg} as ``Lexical Elements Activation Query''.

\paragraph{\textbf{YAGO Ontology triggering}}
Lexical grounding from WordNet is even more relevant due to the possibility to retrieve entities from YAGO ontology aligned as \texttt{owl:sameAs} WordNet synsets. This query is shown in Fig. \ref{fig:populate_kg} as ``YAGO Ontology query''.

\paragraph{\textbf{Close Match triggering}}
Finally, a broader type of triggers retrieval is done by considering entities having a \texttt{skos:closeMatch} to FrameNet frames declared as triggers of some value/violation.
This query allows to assert as triggers mainly entities from WordNet, VerbNet, BabelNet and PropBank, and it is shown in the workflow in Fig. \ref{fig:populate_kg} as ``CloseMatch query''.

After creating a knowledge base of semantic triggers for each and any of the more than 300 Folk Values, the next step is to use this knowledge base to perform values extraction tasks, described in Section \ref{sec:eval}.
Note that, in this work we perform values extraction from natural language, but thanks to the variety and multi-modality of value triggers, coming from many different resources, the ValueNet ontology, and this newly introduced module can be used to perform value extraction from any kind of annotated and/or aligned dataset on the web.

\section{That's All Folks}
\label{sec:TAF}

As stated in the Introduction, the FOLK and TAF modules stem from the awareness that existing value theories and morality-attribution frameworks habe a limitation: they aim at describing high level notions and concepts, as exposed in Sec. \ref{sec:theoretical_grounding}, with the purpose to develop a pan-cultural (universally shared) model, however they miss value terns dealing with everyday life and daily conversation situations. 
In the well established definition proposed in the Introduction, Values are defined as ``principles, beliefs, and attitudes that guide our behavior and shape our interactions with others''. Therefore, we present here a commonsense folk-driven ontology, theoretically complementary to BHV and MFT, which aims at representing and reasoning about generalised commonsense values. 
In order to gather as many values as possible, we adopted a bottom-up approach as follows:
\begin{enumerate}
    \item Scrape the web to gather all the main lists of so-called ``values'' being them qualified as ``cultural'', ``inner'', ``personal'', ``core'', etc., collecting more than 350 potential Folk Values, mainly from 7 different URLs\footnote{All the URLs of the online resources used to gather the complete list of Folk Values are available here: \url{https://github.com/StenDoipanni/ValueNet/blob/main/ThatsAllFolks/URLs.txt}}; these FV were then introduced in the TAF ontology, keeping track of their provenance via the prov-o property \texttt{prov:wasAttributedTo} pointing at the original URLs scraped;
    \item Manually analyse the list, in order to filter the granularity of detail, dedupe entities pointing at the very same semantic space (e.g. \texttt{folk:Winning} and \texttt{folk:Victory}) and determine a taxnomy among them;
    \item Treat those values as frames, therefore represent them as classes of situations for which it is possible to individuate roles, lexical triggers and factual entities that, in their semantics, point at a FV related occurrences of a certain situation.
    \item Align them, where possible, to BHV and MFT values in ValueNet;
    \item Populate the knowledge graphs with triggers from FrameNet, WordNet, VerbNet, YAGO, Wikidata, BabelNet, etc.
\end{enumerate}

The final modules includes more than 300 folk-values, formalised as frames, aligned to FrameNet and whose knowledge graphs of triggers span from lexical resources like WikiData, VerbNet, and WordNet, to factual ones like DBpedia, Umbel, YAGO, as well as ConceptNet, Propbank, and of course FrameNet.

Some of the Folk Values worth to be mentioned, namely those that are completely ignored by main theories, but still are recognised as determinant in everyday life can be: \texttt{folk:Intelligence} for which many subClasses of value situations have been retrieved, such as \texttt{folk:Brilliance}, different from \texttt{folk:Cleverness} for its more punctual extension in time (not yet modeled, but mentioned in the next Sections as future works); \texttt{folk:Learning}, described as the desire to learn more than what is already known, and \texttt{folk:Wisdom} often attributed (in a biased way) to entire subsets of individuals.

To test this new modules of the ValueNet ontology we performed an automatic value detection, using a frame-based method, as explained in the next Section.

\section{Resource Validation}
\label{sec:eval}

Although it is a difficult task to validate a resource of this kind, since its meaning is to be able to catch those shades that are (sometimes for good reasons) excluded by pan-cultural theories, and for which therefore (i) subjectivity is intrinsic to the matter and (ii) there is no data nor baseline to adopt, we conducted experiments to both test the coverage increment, and the inference power.

In order to do so we reused the FRED \cite{gangemi2017semantic} tool. FRED can be described as a ``situation analyzer'', in fact, it is a system for hybrid knowledge extraction from natural language, based on both statistical and rule-based components, which generates RDF/OWL knowledge graphs, embedding entity linking, word-sense disambiguation, and frame/semantic role detection.

Furthermore, to test the resource we selected from the Moral Foundation Reddit Corpus \cite{trager2022moral} a subset of the first 1k sentences.
Each sentence is passed to FRED to generate a kowledge graph, and then values are extracted using an ad hoc  developed frame-based automatic detector for the experiment\footnote{It is possible to test a beta version of the value detector available online here: \newline \url{http://framester.istc.cnr.it/semanticdetection/values}, this version uses only triggers from MFT module, but the full frame-based value detector tool will be released with the camera ready.}.

The frame-based detector operates in this way: it produces the graph with the FRED tool, already aligned and disambiguated on Framester already mentioned semantic web resources, and then, for each of these nodes, it performs a SPARQL query to the Framester endpoint, looking for a triple declaring the activation of some value from one of the modules in ValueNet, if the query gives positive results, it attaches the triple to the original graph.

\begin{table}[h]
\setlength{\tabcolsep}{9pt}
\centering
\caption{Total amount of sentences per annotator, agreement  }
\label{tab:eval}
\begin{tabular}{@{}| l | c | c | c | c | c |@{}}
\toprule
Annotators &  Tot  &  Tot-NC  &  \makecell{Agree / \\ TOT } &  \makecell{Agree+TM / \\ Tot} &  \makecell{Agree+TM/ \\ Tot-NC}    \\ \midrule
A00 & 157 & 63 & 52 & 62 & 34 \\
A01 & 137 & 136 & 53 & 60 & 60 \\
A02 & 185 & 180 & 65 & 75 & 75 \\
A03 & 302 & 296 & 122 & 130 & 130 \\
A04 & 163 & 163 & 6 & 63 & 63 \\
\bottomrule
\end{tabular}
\end{table}

Table \ref{tab:corpus} shows some quantitative data: out of the 1k subset, shown in column ``MFRC'', in 944 cases, column ``FRED Subset'', the FRED tool successfully generated a knowledge graph. The reasons for the missing sentences not producing any knowledge graph could be due to many different problems: irregular syntax, brevity of sentences, use of abbreviation or not-recognised slang (e.g. ``imho'' for ``in my humble opinion'', etc.), or even problems in character encoding.
Table \ref{tab:eval} shows some results of the analysis, in fact, the 944 sentences are not the set of unique sentences, this would in fact be of 306 sentences, since the original corpus was realised using 5 annotators, granting each sentence to be labeled by at least 3 annotators. Therefore, each sentence annotated by each annotator is considered a token \textit{per se}: in Table \ref{tab:eval} column ``Tot'' shows the amount of sentences annotated by each annotator in the considered subset. The original dataset included also a confidence score, expressed as ``Confident'', ``Somewhat Confident'', and ``Not Confident''. 
Column ``Tot-NC'' shows the amount of sentences per each annotator if we exclude those for which the confidence score is equal to ``Not Confident''.
Taking into account that this is a subset, it is still worth to note the uncertainty and intrinsic subjectivity of value annotation task: as Table \ref{tab:eval} shows, A00 seems to be not confident almost half of the time, while A03 expresses confidence in 98\% of the annotations.

\begin{table}[h]
\setlength{\tabcolsep}{8pt}
\centering
\caption{Total sentences and Original Annotation vs Frame-based Detected Values}
\label{tab:corpus}
\begin{tabular}{@{}| l | c | c | c | c | c | @{}}
\toprule
MFRC &  \makecell{FRED \\ Subset}  &  \makecell{MFRC \\ Annotation } & TM & NM & \makecell{Detected}  \\ \midrule
1000 & 944 & 228 / 1000 & 153 & 563 & 855 / 944  \\
\bottomrule
\end{tabular}
\end{table}

\begin{figure}

\label{fig:mfrc_original}
  \includegraphics[scale=0.5]{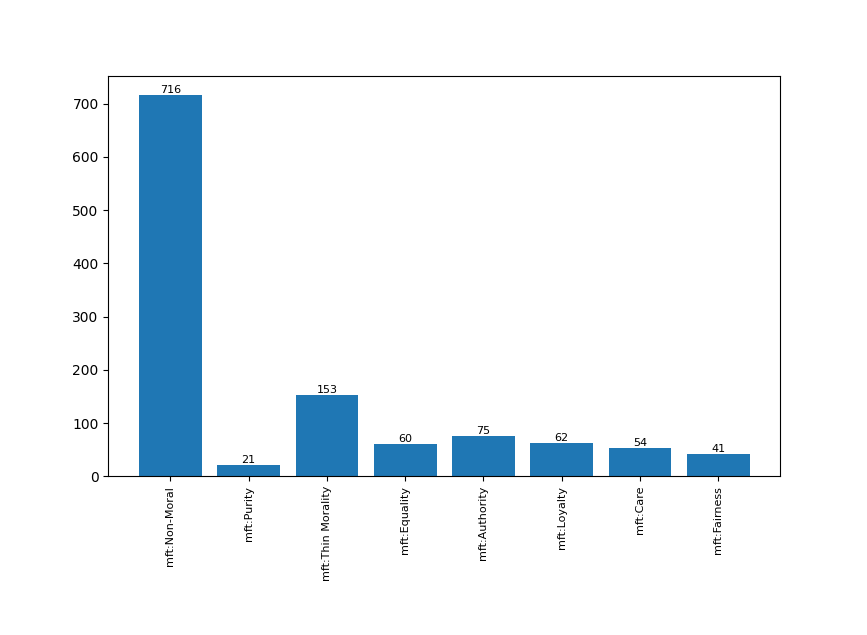}
  \caption{Original annotation of the MFRC dataset.}

\end{figure}

\begin{figure}
\label{fig:mfrc_detected}
  \includegraphics[scale=0.4]{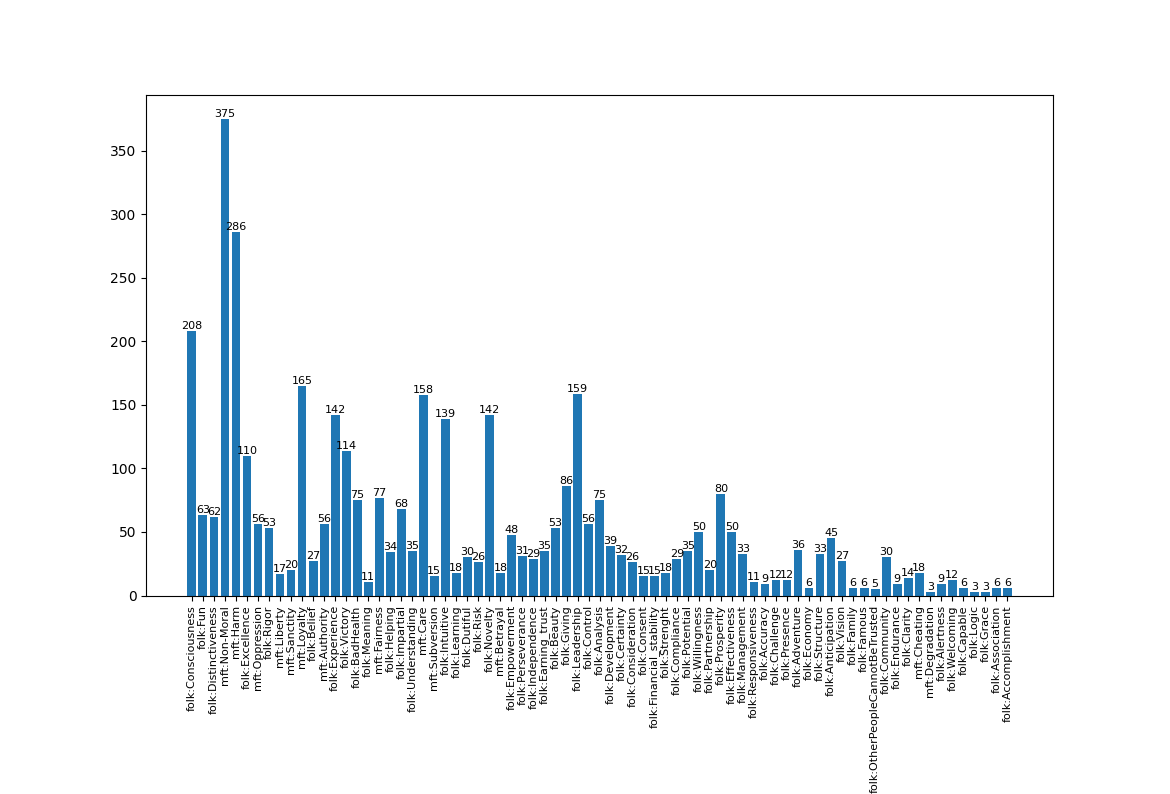}
  \caption{Folk Values and MFT Values detected.}
\end{figure}

A surprising datum is shown both in Fig. \ref{fig:mfrc_original} and Table \ref{tab:corpus}: out of 944 sentences, in the original annotation only 228 of them were tagged with at least one MFT value, as shown in Table \ref{tab:corpus}, column ``MFRC Annotation''. In fact, in 716 cases the sentence was labeled as ``Non-Moral'' (563 in column ``NM'') or ``Thin Morality'' (153 in column ``TM''), which, for the purpose of our analysis, we could paraphrase as: in 563 cases the MFT values were not retrieved in the sentence, while in 153 cases the annotator recognised that there was some sort of morality, but the MFT values were not enough to catch that specific / more subtle / cultural-dependent morality shade that the annotator still was claiming to be there.
Column ``Detected'' in Table \ref{tab:corpus} shows instead that, out of 944 total cases, in 855 of them at least one Folk Value is detected, 635 of which overlaps the subset of 716 cases for which zero or not-specified morality was originally indicated, Fig. \ref{fig:mfrc_detected} shows the detailed amount of activation occurrences per each value, both from FOLK and MFT. This alone, as clearly shown in Fig. \ref{fig:mfrc_detected}, means a significant increment of the semantic information about latent moral content.
To proceed with a more qualitative analysis: it seems that the subset that it is meaningful to be analysed here is the one composed by sentences that were labeled as having a ``Thin Morality'', for which we can assume that the MFT values were not sufficient.

Let's therefore take an example, in order to show the automatic inferences allowed by the graph structure and the semantic dependencies.
Graph n° 357 is generated out of the following sentence\footnote{All the graphs generated by FRED and labeled with the value detector are available on the TAF repository: \url{https://github.com/StenDoipanni/ValueNet/tree/main/ThatsAllFolks/MFRC\_1k\_graphs}}:
\begin{quote}
        \textit{And however flawed or \underline{dishonest} Macron may be.....it is a far greater act of dishonesty to \underline{steal} his data and \underline{expose} it, hoping to change the course of a \underline{national} election for the purpose of an outside group. That is far far more \underline{dangerous} than voting for one flawed man.}
\end{quote}
   
The sentence above is labeled by Annotator00 and Annotator03 as \texttt{Thin Morality} while it is \texttt{Non-moral} for Annotator01.
The frame-based detector annotates this sentence with: \texttt{mft:Loyalty} and \texttt{mft:Betrayal} from MFT, and \texttt{folk:Rigor}, \texttt{folk:Learning} and \texttt{folk:Risk} from the TAF module.
The full graph is not shown here for visualization reasons but it is available on the GitHub\footnote{The full graph is available here: \newline \url{https://github.com/StenDoipanni/ValueNet/blob/main/ThatsAllFolks/eswc\_thin\_folk.png}}.
What is relevant is that, with the whole structure of semantic dependencies, it is possible to have the co-occurrence of some apparently conflictual tags, e.g. in this case the activation of both \texttt{mft:Loyalty} and \texttt{mft:Betrayal}.
Analysing the graph, in fact, it is possible to retrieve the exact topology of activation, keeping track of the role of the value-trigger in the value situation.
In this case \texttt{mft:Loyalty} stems from the \texttt{fs:Candidness} FrameNet frame, evoked by the ``dishonest'' lexical unit, and the WordNet synset for the adjective \texttt{wn:national-adjective-1}. The \texttt{mft:Betrayal} value is instead triggered by the \texttt{fs:RevealSecret} FrameNet frame, which is evoked by the VerbNet entity \texttt{vb:Expose\_48012000}. As for folk values, \texttt{folk:Learning} is unfortunately activated by an incorrect disambiguation of the lexical unit ``course'', creating a bit of noise; \texttt{folk:Rigor} is instead triggered by the \texttt{fs:Law} FrameNet frame, evoked by the segment ``act of dishonesty'' and \texttt{folk:Risk} is triggered by the \texttt{wn:dangerous-adjective-1} WordNet synset and the \texttt{fs:RiskySituation} frame.

Furthermore, via some graph-pattern heuristics, combining the VerbNet roles and the affect stance of the verbs (available in the Framester resource), it is furthermore possible to retrieve, via SPARQL query, knowledge about the subject of the VerbNet entity \texttt{vn:Steal\_10050000}, which is modeled as having a ``socially reprehensible'' negative value on the Agent (in our case the \textit{dishonesty} node).

We are therefore able to extract much and much more varied knowledge about the distribution of the sentence's value load, expliciting latent moral content, and offering an explainability that in a flat table would seem an inconsistency.

\section{Conclusions and Future Work}
\label{sec:conclusion}

We presented a resource thst represents and enables the detection of Folk Values, namely values as intended in commonsense knowledge during everyday interactions. Its purpose is to complement existing  theories, and to provide a method for identifying the values that are not contemplated those theories, but nonetheless play a key role shape the social structure, cultural biases, and personal beliefs.
The evaluation performed show meaningful results both in the quantity and quality of data retrieved, with a significant increment of detection of MFT values, compared to the original MFRC resource, as well as the introduction of labeling with entities from the FOLK ontology, a novel resource introducing more than 300 new values from commonsense knowledge.
Future work includes maintaining dynamic evolution and data quality of the resource, and specification of relations among Folk Values, which are already aligned to MFT and BHV modules in the ValueNet ontology.

A further development is the introduction of emotion detection on the graph, to elaborate more complex and complete patterns.

\section*{Acknowledgment}

This work is funded by the SPICE EU H2020 Project 870811 within the program: SOCIETAL CHALLENGES - Europe In A Changing World - Inclusive, Innovative And Reflective Societies and by the H2020 Project TAILOR: Foundations of Trustworthy AI – Integrating Reasoning, Learning and Optimization -- EC Grant Agreement number 952215

%
%
%
\bibliographystyle{splncs04}
\bibliography{bib.bib}

\end{document}